# Quantum key distribution system clocked at 2 GHz

**Karen J. Gordon, Veronica Fernandez, Gerald S. Buller**
*School of Engineering and Physical Sciences, Heriot-Watt University, Edinburgh, UK, EH14 4AS*
*k.j.gordon@hw.ac.uk*

**Ivan Rech, Sergio D. Cova**
*Dipartimento Elettronica e Informazione, Politecnico di Milano, 20133, Milano, Italia*

**Paul D. Townsend**
*PhotonicsSystems Group, Department of Physics, University College Cork, Cork, Ireland*

*http://www.phy.hw.ac.uk/resrev/photoncounting/index.html*

**Abstract**: An improved quantum key distribution test system operating at clock rates of up to 2GHz using a specially adapted commercially available silicon single photon avalanche diode is presented. The use of improved detectors has improved the fibre-based test system performance in terms of transmission distance and quantum bit error rate.

**Introduction:** Quantum key distribution (QKD) enables two users, Alice and Bob, to share a verifiably secure encryption key, guaranteed by the laws of quantum mechanics [1].

Since its first experimental implementation in 1992, the growth towards practical applications has been rapid, both in the use of optical fibres as the transmission medium [2,3], and in free-space transmission systems [4,5]. Whilst much experimental effort has been made to increase the transmission span of such point-to-point systems (currently demonstrated at up to ~120km [3]), the key exchange rate still remains low in such systems

– typically <1kbits$^{-1}$. This is particularly true in the case of 1.55μm wavelength QKD systems due to count rate limitations imposed by the deleterious effects of the afterpulsing phenomenon evident in the cooled InGaAs/InP single photon avalanche diode (SPAD) detectors used. However, a different approach to increase the potential key exchange rates, utilising the mature technology of Si SPAD detectors in conjunction with standard telecommunication fibres, has been exploited by these authors at gigahertz clock rates [6].

In this Letter, we present a modification of the QKD system to include an electronically enhanced commercially available silicon single photon counting module (SPCM), allowing faster clock rates to be employed. We show that the use of the enhanced module in the QKD system enables the capability of operating up to 2GHz clock rates. The system was characterised in terms of quantum bit error rate (QBER), as discussed previously in [6].

**Description of the system:** The gigahertz QKD system [6] utilised the B92 protocol [1], which requires only two non-orthogonal states. This protocol was achieved by using two linear polarisation states, 45° apart with respect to each other. Two vertical-cavity surface-emitting lasers (VCSEL's) were used at Alice as the sources of the two linearly polarised encoding states. To reduce the probability of more than one photon per pulse, both VCSEL outputs were attenuated to achieve an average number of approximately 0.1 photons per pulse. The VCSEL's had an emission wavelength of ~850nm. The system was clocked optically by multiplexing 1.3μm wavelength synchronisation pulses with the 850nm wavelength encoded photons. These pulses were detected by a linear gain Ge avalanche photodiode (APD), whose output was directed to the synchronisation input of the photon-counting acquisition card. The encoded photons were detected using commercially available silicon Perkin Elmer SPCM-AQR single photon detectors.

**Experiment:** We show a significant improvement in experimental data in QBER by comparing data taken using a standard Perkin Elmer SPCM-AQR photon detector and a specially adapted module of the same type. This device was adapted at Politecnico di Milano, Italy by inserting a new pulse-processing circuit designed for improving the photon timing performance as described in reference [7].

Three main factors cause the QBER to increase with increasing clock frequency: (1) broadening and patterning of the of the VCSEL output pulses due to the limited bandwidth of the laser and associated drive electronics (2) pulse broadening due to dispersion in the fibre; and (3) the timing jitter of the single photon detectors at the receiver Bob. The most significant contributor to QBER is the detector timing jitter.

The temporal response of the SPCM module was improved both in terms of timing jitter (see Fig.1) and centroid shift of the time distribution associated with high count rates (typically above $0.5 Mcounts^{-1}$). At low counting rates the original module prior to enhancement had a full width at half maximum (FWHM) jitter of ~570ps. After adaptation this device exhibits a FWHM jitter of ~370ps. More significantly the modified device almost completely eliminates the additional temporal broadening observed in the original module at high incident count rates (of greater than $0.5 Mcounts^{-1}$), and the accompanying centroid time shift. For example, at an incident count rate of $2 Mcounts^{-1}$ the modified device exhibits a jitter of ~450ps (FWHM), compared with ~950ps jitter prior to modification. Temporal broadening of the single photon detector has been shown to limit *the* performance of the QKD system [6] since at clock frequencies between 1 and 2GHz and short fibre lengths the detected count rate can be between 0.5 to 1.5Mcounts$^{-1}$. The reduction in the centroid shift does not directly improve the QBER, however it does allow the data collection window to stay fixed with respect to the synchronisation pulse [6].

Fig. 2 shows the improvement in QBER over a range of high clock frequencies from 1GHz to 2GHz. Comparing the standard SPCM module and the enhanced module for a fixed fibre length of 6.55km the QBER significantly drops below 10% between 1 and 2GHz. At a clock rate of 2GHz the QBER halves from the prohibitively high figure of ~18% to ~7%, the lower value being regarded as being secure from eavesdropping attacks [8].

The significant improvement at a clock frequency of 2GHz is further illustrated in Fig. 3. Fig. 3 shows QBER versus fibre length at a fixed clock frequency of 2GHz. It is clear that the QBER has dropped to a practical level due to the electronic enhancement in the temporal response of the SPCM module. The slight increase in QBER at short distances for the standard detector is due to the temporal broadening at high-count rates. Furthermore, these results indicate that use of single photon detectors with a faster temporal response [9] than the SPCM modules currently used in the QKD system offer the potential benefits of lower QBER and the consequent advantages of longer distance key distribution and/or higher key exchange rates.

Additionally, at a clock frequency of 2GHz for a fixed fibre length of 6.55km the estimated net bit rate after error correction and privacy amplification improved from zero to the order of 20kbits$^{-1}$ due to the decrease in QBER.

Conclusion: We have shown that by shortening the temporal response of the single photon detector employed has significantly improved the performance of the quantum key distribution system at clock frequencies greater than 1GHz. The system has been improved in terms of increasing the workable clock frequency range from 1GHz to 2GHz, but also the results at higher frequencies have improved in terms of transmission distance. For a fixed fibre length of 6.55km and clock rate of 2GHz the QBER was improved from 17.8% to 6.6%. Further improvements in transmitter and detector timing resolution will further improve system performance, for example the introduction of faster shallow junction single photon avalanche diode detectors [9] and higher bandwidth driving electronics and VCSEL's.

**Figure 1:** *Timing jitter full width at half maximum of both SPAD modules.*

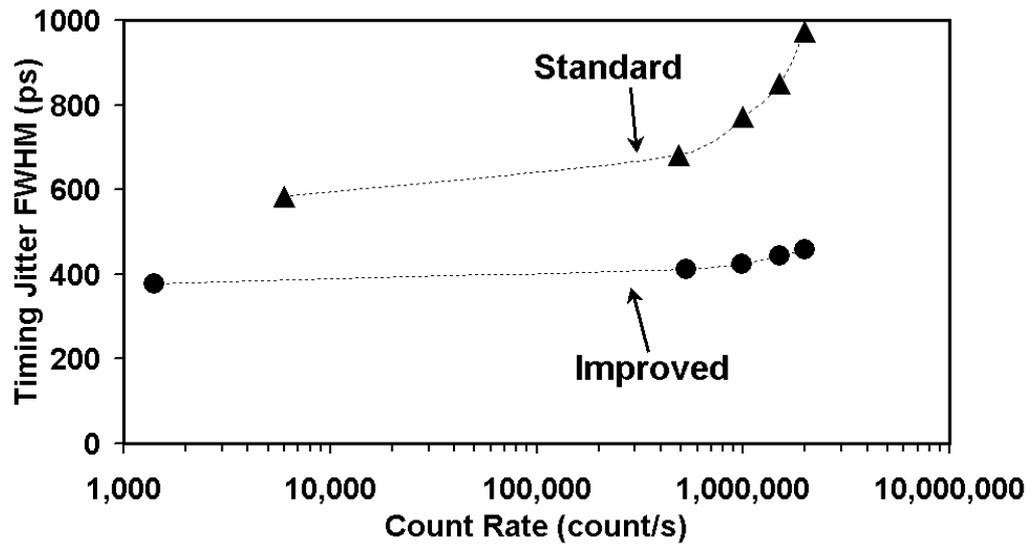

**Figure 2:** *QBER versus QKD system clock frequency at fixed fibre distance of 6.55 km of standard telecommunications fibre*

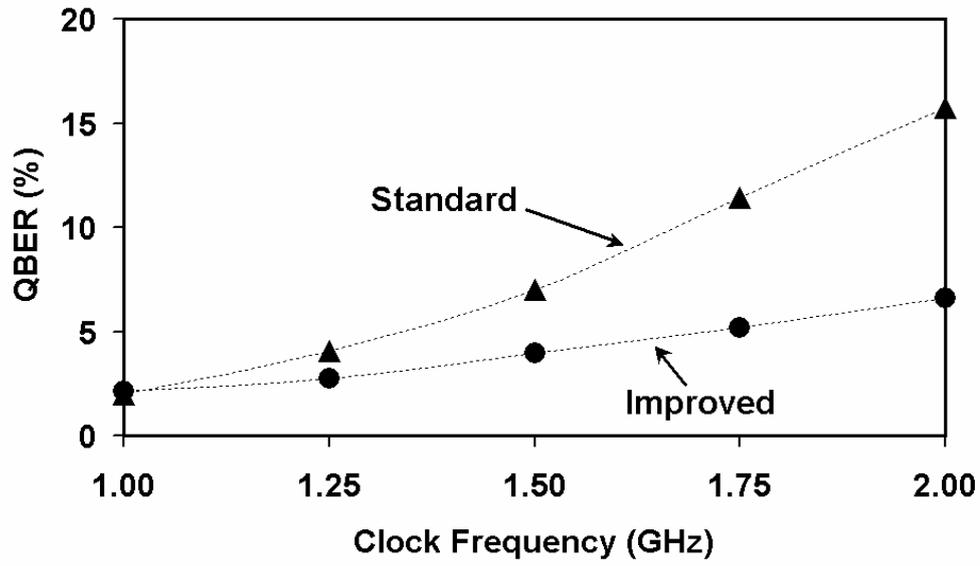

**Figure 3**: *QBER versus fibre distance of at a clock frequency of 2GHz. The points filled in black are taken with the full fibre transmission distance. The white points were measured using optical attenuation to simulate the given distances.*

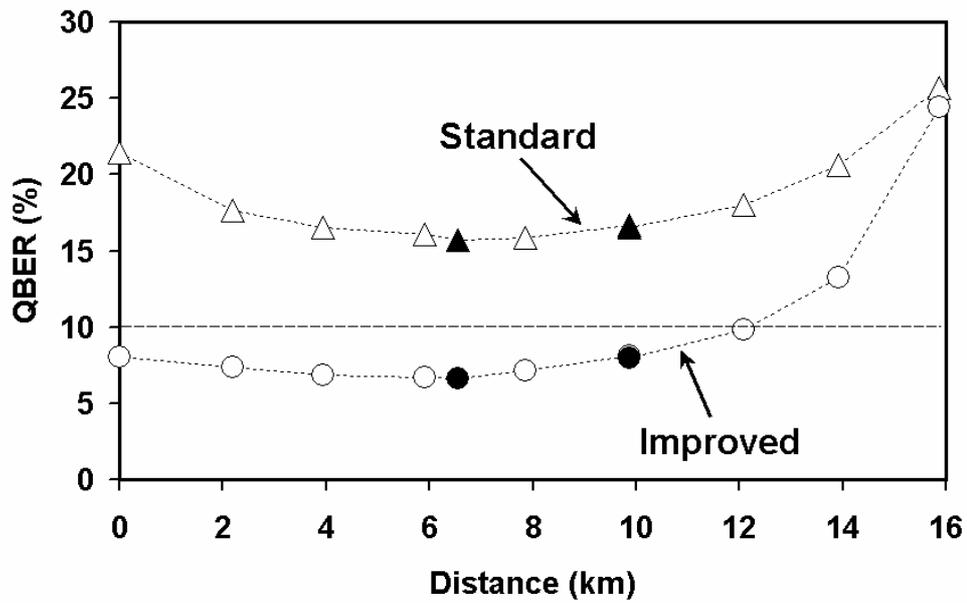